\documentclass[final,5p,times,twocolumn]{elsarticle}
\biboptions{numbers,sort&compress}

\newif\ifdraftmode
\draftmodetrue

\newif\ifshowsplattext

\usepackage{etex}
\usepackage{subfigure}

\usepackage{pslatex}
\usepackage{microtype}

\usepackage{subfigure}
\usepackage{listings}
\usepackage{colortbl}
\usepackage{amsfonts}
\usepackage{amsmath}
\usepackage{pifont} 
\usepackage{proof} 
\usepackage{mathpartir}
\usepackage{fancyvrb}\fvset{fontsize=\small}
\usepackage{xspace}
\usepackage{balance}
\usepackage{multirow}
\usepackage[T1]{fontenc}
\usepackage{relsize}
\usepackage{tikz}
\usepackage{tikz-qtree}
\usepackage{longtable}
\usepackage{booktabs}
\usetikzlibrary{fit,shapes,calc,positioning,shadows,arrows,shapes,backgrounds,
decorations.markings,fadings}
\usepackage{wrapfig}
\usepackage{dsfont}
\usepackage{natbib}

\definecolor{bgBlock}{rgb}{0.22,0.15,0.49}
\definecolor{bgBlockAlert}{rgb}{0.99,0.84,0.31}
\definecolor{fgBlockAlert}{rgb}{0.22,0.15,0.49}
\definecolor{fgBlock}{rgb}{0.99,0.84,0.31}

\renewcommand{\ttdefault}{pcr}
\usepackage{color}
\definecolor{darkred}{rgb}{0.5,0,0}
\definecolor{darkgreen}{rgb}{0,0.5,0}
\definecolor{darkblue}{rgb}{0,0,0.5}

\usepackage[bookmarks=false]{hyperref}
\hypersetup{ colorlinks,
  linkcolor=darkblue,
  filecolor=darkgreen,
  urlcolor=darkred,
  citecolor=darkblue }
\usepackage{url}

\newcommand{\Comment}[1]{}
\ifdraftmode
  \newcommand{\Fix}[1]{\textbf{[[}{\color{red} #1}\textbf{]]}}
  \newcommand{\marcelo}[1]{\textbf{[[Marcelo: }{\color{magenta} #1}\textbf{]]}}
  
  \newcommand{\note}[1]{\todo[inline,color=red!30,caption={}]{#1}}
\else
  \newcommand{\Fix}[1]{\relax}
  \newcommand{\marcelo}[1]{\relax}
  \newcommand{\note}[1]{\relax}
\fi

\newcommand{\ie}[1]{\emph{i.e.}}
\newcommand{\eg}[1]{\emph{e.g.}}
\newcommand{\aka}[1]{\emph{aka}}
\newcommand{\etal}{\emph{et al.\/}}
\newcommand{\tnameName}{EvoSPLat}
\newcommand{\tname}{\CodeIn{\tnameName}}
\newcommand{\RegressionSPLKorat}{\tname}
\newcommand{\evopairwise}{\CodeIn{(Evo)Pair-wise}}
\newcommand{\evosixwise}{\CodeIn{(Evo)Six-wise}}

\newcommand{\rcs}{RCS}
\newcommand{\rts}{RTS}
\newcommand{\gcc}{GCC}
\newcommand{\spl}{SPL}
\newcommand{\splat}{\SPLKorat{}}

\newcommand{\SPLKoratName}{SPLat}
\newcommand{\SPLKorat}{\CodeIn{\SPLKoratName}}
\newcommand{\relevant}{relevant}

\newcommand{\Space}[1]{}

\newcommand{\options}{variables}
\newcommand{\satisfiable}{satisfiable}

\newcommand{\figref}[1]{Figure~\ref{fig:#1}}
\newcommand{\definit}[1]{\textcolor{blue}{\emph{#1}}}
\newcommand{\red}[1]{\textcolor{red}{\emph{#1}}}

\newcommand{\numsubjects}{12}

\let\Itemize =\itemize    
\let\Enumerate =\enumerate
\let\Description =\description
\def\Nospacing{\itemsep=0pt\topsep=0pt\partopsep=0pt\parskip=0pt\parsep=0pt}
\renewenvironment{itemize}{\Itemize\Nospacing}{\endlist}
\renewenvironment{enumerate}{\Enumerate\Nospacing}{\endlist}
\renewenvironment{description}{\Description\Nospacing}{\endlist}
\makeatletter
\def\@copyrightspace{\enlargethispage{-10pt}\relax}
\makeatother

\def\z{\phantom{0}}

\newcommand{\codesize}{\Comment{\smaller}}
\newcommand{\CodeIn}[1]{\codeid{#1}}
\def\|#1|{\mathid{#1}}
\newcommand{\mathid}[1]{\ensuremath{\mathit{#1}}}
\def\<#1>{\codeid{#1}}
\newcommand{\codeid}[1]{\ifmmode{\mbox{\codesize\ttfamily{#1}}}\else{\codesize\ttfamily #1}\fi}

\newcommand{\cmark}{\ding{51}}%
\newcommand{\xmark}{\color{red}\ding{55}}%

\newcommand{\twise}{$t$-$\mathit{wise}$}
\newcommand{\gccversion}{4.8}

\begin{document}

\title{Time-Space Efficient Regression Testing for Configurable Systems}



 

\author[uepb]{Sabrina Souto}
\address[uepb]{Universidade Estadual da Para\'iba, Campina Grande, PB, Brazil}
\ead{sfs@cin.ufpe.br}

\author[ufpe]{Marcelo d'Amorim} 
\address[ufpe]{Universidade Federal de Pernambuco, Recife, PE, Brazil}
\ead{damorim@cin.ufpe.br}
 
\begin{abstract}

Configurable systems are those that can be adapted from a set of options. 
They are prevalent and testing them is important and challenging. Existing 
approaches for testing configurable systems are either unsound (\ie{}, they 
can miss fault-revealing configurations) or do not scale.

This paper proposes \tname{}, a regression testing technique for configurable 
systems. \tname{} builds on our previously-developed technique, \SPLKorat{}, which 
explores all dynamically reachable configurations from a test. \tname{} is 
tuned for two scenarios of use in regression testing: Regression Configuration 
Selection (\rcs{}) and Regression Test Selection (\rts{}). \tname{} for \rcs{} prunes 
configurations (not tests) that are not impacted by changes whereas \tname{} 
for \rts{} prunes tests (not configurations) which are not impacted by changes. 
Handling both scenarios in the context of evolution is important.

Experimental results show that \tname{} is promising. We observed a substantial 
reduction in time ($\sim$22\%) and in the number of configurations ($\sim$45\%) for 
configurable Java programs. In a case study on a large real-world configurable 
system (GCC), \tname{} reduced $\sim$35\% of the running time. Comparing \tname{} 
with sampling techniques, 2-wise was the most efficient technique, but it missed 
two bugs whereas \tname{} detected all bugs four times faster than 6-wise, on average.

\end{abstract}

\maketitle

\section{Introduction}

Configurable systems are those that can be adapted through input
options, reflected in code in the form of variations.  Large software
systems often provide some level of configurability to users or to
developers.  The intuition is that the ability to reason about these
variations facilitates maintainability and reduces
time-to-market~\citep{gccoptions}.  Configurable systems are prevalent
and the amount of variation they offer can be very
high~\citep{zhang-ernst-icse2013,zhang-ernst-icse2014}.  Examples of
configurable systems include the Firefox web
browser~\citep{firefox-web}, the Linux
kernel~\citep{linux-kernel-web}, the \gcc{} compiler
infrastructure~\citep{gcc-web}, and the deals-recommendation web
service Groupon~\citep{groupon-web}.

Testing configurable systems is an important problem that continues to
attract a lot of attention from the research
community~\citep{cohen-etal-icse2003,Qu-issta2008,emine-etal-issta2011,kim-etal-isrre2012,kim-etal-fse2013,nguyen-etal-icse2014,sabrina-etal-splc15,medeiros-etal-icse2016,souto-etal-icse2017}.
Conceptually, the large space of possible configurations makes the
introduction of errors easier and testing for those errors more
challenging.  For the sake of cost, it is not uncommon practice to
test the system against a single configuration\footnote{Such
  configuration is often referred to as the ``default'' configuration
  and includes the features which are more typical to build the
  system.}~\citep{sabrina-etal-splc15,medeiros-etal-icse2016}.  For
instance, Groupon~\citep{kim-etal-fse2013} and
\gcc{}~\citep{sabrina-etal-splc15} follow this practice.  At another
extreme, exhaustively testing all configurations is unacceptably
expensive.  Large software systems typically offer hundreds of
configuration options, leading to a combinatorial blowup in the number
of possible configurations to test.

This work focuses on the problem of \emph{regression testing
  configurable systems}.  Regression testing is an important
quality-assurance activity realized during software evolution to
reduce the chances of defects escaping to production.  It consists of
repeatedly executing a test suite and monitoring its effects with the
goal of anticipating the observation of errors.  Regression testing is
notoriously expensive and continues to receive huge attention from
researchers and
practitioners~\citep{rothermel-etal-tse2001,engstrom-etal-infosot2010,yoo-harman-stvr2012,bell-kaiser-icse2014,elbaum-etal-fse2014,gligoric-etal-issta2015,hilton-etal-ase2016}.
In the context of configurable systems, regression testing becomes
even more expensive as each test needs to be executed against several
different configurations.  Despite the interest in regression testing
from the research community and the prevalence of configurable systems
in practice, research on this problem, which intersects two
very-active areas, is surprisingly scarce.

To solve this problem in general a technique intuitively needs to
identify the impact of evolutionary changes on the execution of each
test.  Unfortunately, statically finding a sound yet small set of
relevant configurations for each test is challenging~--~computing
accurate static approximations of dynamic impact sets precisely and
efficiently for non-configurable system is already a challenging
problem~\citep{law-rothermel-icse2003,apiwa-etal-icse05}; adding the
configuration dimension does not make the problem any
simpler~\citep{spllift,lillack-ase2014,angerer-ase2015}.  Most
previous research on this problem focused on the proposal of
heuristics to find configurations seemingly-related to
changes~\citep{Qu-etal-icsm2007,Qu-issta2008,Qu-etal-icsm2012,Xu-etal-splc2013}. The
strategy scales but can lead to error misses as relevant
configurations can be ignored.


This paper proposes \tname{}, a lightweight dynamic technique for
efficient regression testing of configurable systems.  \tname{} builds
on our previously-developed technique,
\SPLKorat{}~\citep{kim-etal-fse2013}, which explores all
configurations of a system that are dynamically reachable from an
input test.  \tname{} optimizes both time and space to explore
change-impacted configurations. Our approach is tuned for two
scenarios of use: Regression Configuration Selection (\rcs{}) and
Regression Test Selection (\rts{}).  \tname{} for \rcs{} prunes
configurations (not tests) that are not impacted by changes whereas
\tname{} for \rts{} prunes tests (not configurations) which are not
impacted by changes.  \tname{} uses lightweight static analysis to
observe change impact based on saved information from previous runs.

Both modes of execution are important in regression testing of
configurable systems~(see Section~\ref{sec:modes}).  The \rcs{}
scenario is useful for configurable systems with a wide variety of
tests, including unit and integration tests.  In those cases, each
test typically covers relatively small fractions of the code.  For
scenarios where most tests are system tests, which call an executable
through command-line parameters, \rts{} is likely a better fit.  In
those cases, each test potentially cover most of the code, when
considering all reachable configurations.  Such pattern of tests are
frequently observed in configurable
systems~\cite{abal-etal-ase2014,medeiros-etal-icse2016}.

The contributions of this paper are:
\begin{itemize}
\item A lightweight technique to alleviate cost of systematic testing
  in two important scenarios of use: \rcs{} and \rts{}.
\item An implementation of our technique that is publicly available at
  \url{https://sites.google.com/site/evosplat/}.
\item An empirical evaluation, including software product
  lines~(\spl{}s) and one large program (\gcc{}).  We evaluated
  \tname{} for \rcs{} on 12 \spl{}s and observed significant reduction
  in both time and space.  We also evaluated \tname{} for \rts{} on
  \gcc{}.  Results show that, \tname{} reduces time by 35\% on
  average, compared to \SPLKorat{}.  Compared to sampling techniques,
  namely \twise{}~\citep{nie-leung-acm-surveys2011} with t=2 and 6, we
  observed that \tname{} retained the ability to detect faults as
  6-wise but required much less configurations (time) to achieve that.
  Pairwise, in contrast, was significantly faster compared to \tname{}
  (and 6-wise) but missed 2 out of 5 real \gcc{} bugs we analyzed.
\end{itemize}


\section{Illustrative Example}

This section illustrates \tname{} on a small running example.  

\subsection{Some terminology}
\label{sec:terminology}

We call \definit{configuration variables} (\aka{} feature variables)
those program variables whose purpose in code is to adapt behavior
through variation.  A \definit{configurable system} is one that uses
configuration variables to manage variability in code.  A
\definit{configuration} is an assignment of values to configuration
variables.  In many cases, a \definit{default configuration} exists
(\eg{}, Groupon~\citep{kim-etal-fse2013} and
\gcc{}~\citep{sabrina-etal-splc15}).  Intuitively, the default
configuration works as reference of typical behavior to users and
developers.  In some cases, variation can also be controlled
externally, through \definit{configuration options} defined by users
(not developers).  These options are mapped to configuration variables
in code.  Throughout the text we use the terms variables and options
indistinctly.  It is important to note that a test for a configurable
system takes a configuration as an additional input.


A \definit{feature model (FM)} documents the configuration variables of a 
configurable system and their relationships~\citep{kang-foda-90, feature-oriented-spls-2013}, 
also called \definit{constraints}. A \definit{SAT solver} checks if combination 
of variables or configurations are \definit{legal} or not according to the FM constraints. 
Feature models are optional, because in practice they are not always available.

\subsection{A GPL test}

Figure~\ref{fig:example} shows the test \CodeIn{addEdgeWt} from
GPL~\citep{Lopez-herrejon01astandard}, a library of graph algorithms.
This test builds a graph with 3 connected vertices and checks some
properties on the graph.  For example, the first assertion checks if
the weight 1 can be found through the vertex \CodeIn{v3}.  This should
be the case if the weight list associated with \CodeIn{v3} was
correctly updated with the call to \CodeIn{v3.adjustAdorns(v1, 0)}.
We will refer in the following to three configuration options of GPL:
\CodeIn{WEIGHTED}, \CodeIn{SEARCH}, and \CodeIn{UNDIR}.  For the sake
of illustration, let us assume that the tester uses a default
configuration that has the option \CodeIn{WEIGHTED} set and all other
options unset.  Considering a single configuration, test execution
will cover one distinct path in code where only branches associated
with options which are set will be traversed.  For this pair of test
and configuration, execution \emph{passes}.



  
  

\subsection{\SPLKorat{} - finding configurations for testing}
\label{sec:splat-example}

Given a test for a configurable system,
\SPLKorat{}~\citep{kim-etal-fse2013} determines a set of
configurations on which the test should be run.  \SPLKorat{} finds
configurations as follows.  It executes the test on one configuration,
observes the values of configuration \options, and uses these values
to determine the next configuration the test should be run against.
It repeats this process until it explores all \relevant{}
configurations or until it reaches a specified bound on the number of
configurations. This exploration effectively produces a \definit{decision tree} 
with nodes corresponding to configuration \options{} and edges 
corresponding to the different values these \options{} can hold. 
Leaf nodes indicate whether or not a configuration, associated 
with a complete path in the tree, has been found legal or not.

\setlength{\tabcolsep}{2.5pt}
\begin{figure*}[t!]
\centering
\includegraphics[width=.95\textwidth]{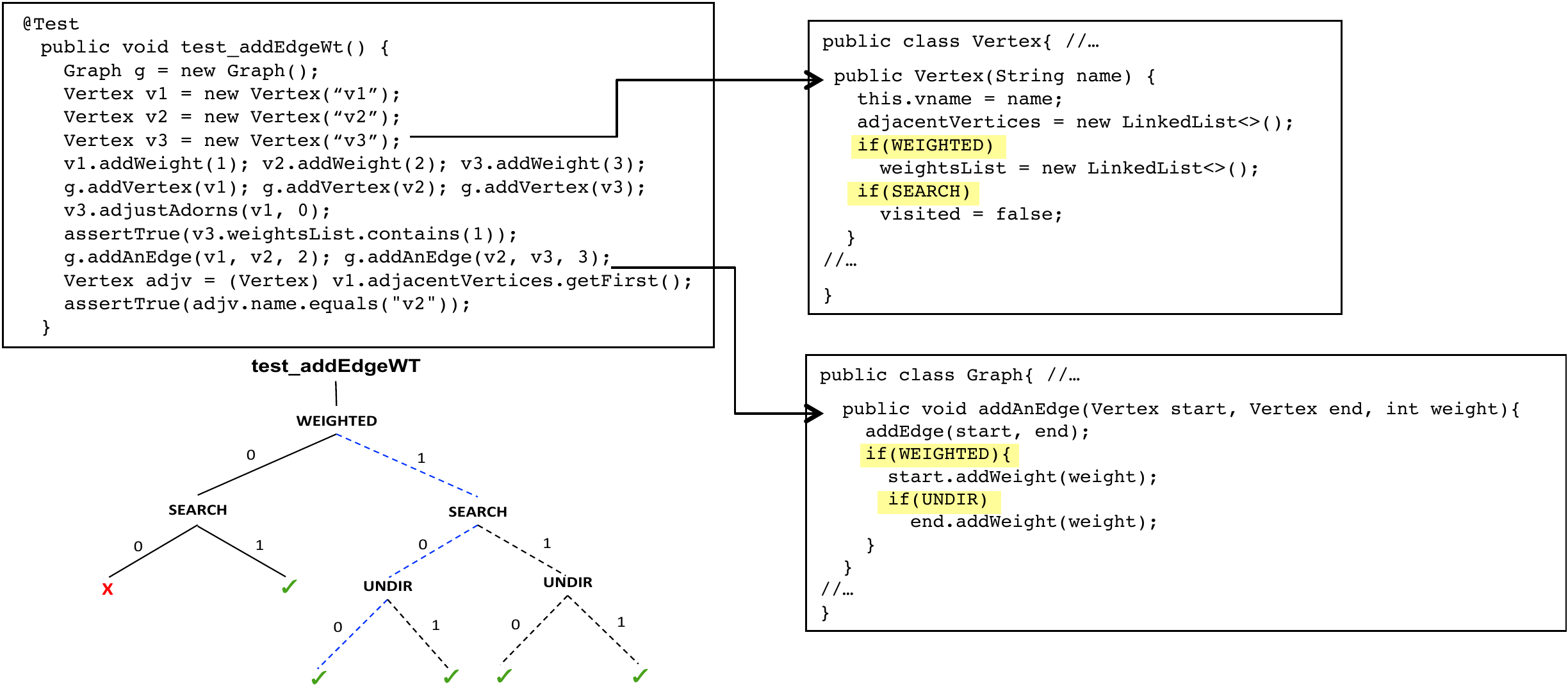}
\vspace{-2ex}
\caption{Test, corresponding decision tree, and fragments of GPL code
  that illustrate how variables are reached through test.}
\label{fig:example}
\end{figure*}

Figure~\ref{fig:example} shows the decision tree obtained with a
complete run of \SPLKorat{} on test \CodeIn{addEdgeWt}.  A path where
variable \CodeIn{WEIGHTED} is set indicates that graph edges have
weights and a path where variable \CodeIn{UNDIR} is set indicates that
the graph is undirected.  \SPLKorat{} starts execution without
assigning concrete values to variables; it assigns values on demand,
when variables are covered.  \SPLKorat{} finds a \emph{sound set} of
configurations to execute the test.  This means that, provided that no
bounds on time or number of configurations exist, \SPLKorat{} cannot
miss fault-revealing configurations for on given input test.
Alternative selection strategies exist (\eg{}, combinatorial
testing~\citep{nie-leung-acm-surveys2011}) but, in contrast to
\SPLKorat{}, they do not provide soundness guarantees.
Section~\ref{sec:related} expands and details related work.

\vspace{1ex}\noindent\textbf{Configuration constraints.}~When the user
provides on input a model that constraints the set of legal
configurations, \SPLKorat{} can safely prune illegal configurations
with the help of a SAT solver.  We emphasize that this model is
optional for \SPLKorat{}. Back to the example, the\Comment{partial}
configuration [\CodeIn{(WEIGHTED,0)},\CodeIn{(SEARCH,0)}] is the first
one that \SPLKorat{} explores.  Note that the \CodeIn{Vertex}
constructor covers variables \CodeIn{WEIGHTED} and \CodeIn{SEARCH},
however variable \CodeIn{UNDIR} is not covered in this example.
\SPLKorat{} discards this configuration as it is illegal according to
the GPL configuration constraints, provided on input for this
case.\Comment{\marcelo{please state which constraint you are talking about.}}
The cross under the leftmost path in the decision tree indicates that.
Then \SPLKorat{} re-executes the test on a new configuration.  It
assigns another value to variable \CodeIn{SEARCH} and re-runs the test
against the configuration [\CodeIn{(WEIGHTED,0)},\CodeIn{(SEARCH,1)}],
which is legal.  This process continues until all reachable
configurations are explored.

\vspace{1ex}\noindent\textbf{Variable types handled.}~\SPLKorat{}
accepts categorical types.  Discretization methods (e.g., domain
partitioning~\cite{ostrand88:category}) can be used to handle
non-categorical domains.



\subsection{\tname{}}
\label{sec:modes}

\SPLKorat{} showed that there are scenarios where exploring all
reachable configurations is feasible.  For example, \SPLKorat{} was
able to explore all dynamically reachable configurations in Groupon
PWA\footnote{Groupon PWA is the name of the codebase that powers the
  \url{groupon.com} website.  See Section 4.2
  from~\citep{kim-etal-fse2013}.} for several test
cases~\citep{kim-etal-fse2013}.  Unfortunately, there are important
scenarios where the number of configurations to explore is still very
high.  \tname{} addresses this inherent scalability issue of
\SPLKorat{}.  In summary, \tname{} leverages information from previous
runs\footnote{\SPLKorat{} is used to bootstrap the process.} to
optimize time and space.

\vspace{1ex}\noindent\textbf{Modes of execution.}  \tname{} provides
two modes of execution: Regression Configuration Selection (\rcs{})
and Regression Test Selection (\rts{}).  In \rcs{} mode, \tname{}
restricts the set of configurations to run on each test; no
restrictions apply to the test set (Section~\ref{sec:evosplat-rct}).
In contrast, in \rts{} mode, \tname{} restricts the set of tests that
will be executed; no restrictions apply to the configurations
associated to each test (Section~\ref{sec:evosplat-rts}).  \tname{}
for \rcs{} benefits the most in a scenario where, for a given test,
distinct configurations cover highly-different sets of functions.  In
that scenario, if a function changes, only few configurations would be
impacted and only few re-executions would be necessary for a given
test.  If, on the contrary, most functions were covered across
reachable configurations, \rcs{} would have little effect as the test
would need to be re-executed in all those configurations. \tname{} for
\rts{} is a better fit for this later scenario; it does not prune
test-reachable configurations but is more lightweight.  The decision
on whether to apply \rcs{} or \rts{} highly depends on the tests of
the application under testing.  Note that \rcs{} generalizes \rts{}.
With \rcs{} one can model a test that should be ignored by associating
an empty set of configurations with that test. \rts{}, however, does
not generalize \rcs{} as it is unable to partially restrict
configurations of tests as \rcs{} does.  Despite this subsumption
relationship, in practice, \rts{} remains important as it enables
optimizations to \tname{}.

\subsection{\tname{} for \rcs{} in a nutshell}
\label{sec:example-evosplat}
\label{sec:space-red}

\tname{} for \rcs{} proceeds as follows.  Initially, \SPLKorat{} is
used as to bootstrap the process.  As result, it produces the entire
decision tree.  When there is a change, a lightweight static analysis
is used to detect which subtrees of the decision tree have been
affected.  Then, a separate execution of \SPLKorat{} runs on each of
these subtrees, potentially inducing changes on the tree structure.
Ideally, when code changes, only a small number of subtrees and paths
will be affected, justifying savings in space.  In a nutshell,
\tname{} for \rcs{} \emph{saves space} by reducing the number of
configurations to explore and it \emph{saves time} by optimizing
constraint solving time.


Back to the example, let's consider the scenario where the developer
changes function \CodeIn{addAnEdge}.  In this case, \tname{} first
applies a lightweight analysis to detect that only the right subtree
of the decision tree was affected by the change and then it spawns a
new execution of \SPLKorat{}, rooted in the configuration
\CodeIn{[(WEIGHTED, 1)]}.  The value of variable \CodeIn{WEIGHTED} is
fixed in this execution.  Hence, \tname{} will only explore four of
the six paths considered in the first execution.  The right decision
subtree shows those paths.  Reduction can be higher or lower depending
on the code changes made during evolution.

\tname{} also capitalizes on the observation that it is possible to
accelerate test execution by caching results of SAT solver queries.
The hypothesis is that more often than not the execution of a test on
\CodeIn{(Evo)SPLat} will produce highly-similar subtrees in
consecutive runs.  To consider the different orderings of accessed
variables, \tname{} stores the constraints in the cache in a canonical
form, which exists for the language of propositions used in this
context.  Considering our scenario of change, \tname{} is able to use
previously-stored results of SAT queries and avoid new calls to the
SAT solver.  Note that, in general, it is not possible to completely
avoid SAT calls as code changes can result in changes in the tree
structure, leading to new constraints.

\subsubsection{\tname{} for \rts{} in a nutshell}

Unfortunately, \SPLKorat{} for \rcs{} will not scale for systems where
most tests reach most configurations; a scenario that occurs often
when testing is performed mostly against a single function, typically
the main function of the system.  In that case, the entire code under
testing is reachable through that function.  This is what happens, for
example, with \gcc{}, where a test consists of a source file augmented
with testing directives indicating what ``features'' of the compiler
should be executed (see Figure~\ref{fig:test_gcc}).  In \gcc{} the
tests execute the same codebase modulo the variations induced from
these input directives.  As result, most functions are called in most
configurations.  For cases like \gcc{}, re-running \SPLKorat{} upon
evolutionary changes is prohibitively expensive.  In a nutshell,
\tname{} for \rts{} identifies which tests have been impacted by
changes and only re-runs \SPLKorat{} on those tests.  \tname{}
maintains a map from tests to functions covered by those tests to
identify impacted tests.  At a high-level this approach to test
selection is similar to that used in
Ekstazi~\cite{gligoric-etal-issta2015}.

\section{Technique}
\label{sec:technique}


This section describes \tname{}, a technique to reduce cost of testing
for configurable systems in the dominant scenario of evolution.
\tname{} offers two modes of execution: \rcs{} and \rts{}.  In the
following, we describe how \tname{} works on each of these two modes.


\subsection{\tname{} for \rcs{}}
\label{sec:evosplat-rct}

\RegressionSPLKorat{} takes as input a feature model, a test, and a
list of methods of interest denoting changes.  It explores relevant
configurations and reports test verdicts on output.  In the following
we present the interface \tname{} uses to check satisfiability of
configurations and then detail the algorithm.

\subsubsection{The Feature Model Interface}

\begin{figure}[t!]
\vspace{2ex}
\normalsize
\begin{lstlisting}[xleftmargin=2pt,framexleftmargin=-1pt,framexrightmargin=-3pt,framexbottommargin=1pt]
class FeatureVar {...}
class VarAssign {... Map<FeatureVar, boolean> map; ...}
interface FeatureModel {
 Set<VarAssign> getValid(VarAssign a);
 boolean isSatisfiable(VarAssign a);
 boolean isMandatory(FeatureVar v);
 boolean getMandatoryValue(FeatureVar v);
}
\end{lstlisting}
\normalsize
\caption{Feature Model Interface}
\label{fig:fmInterface}
\end{figure}

\figref{fmInterface} shows the classes and interfaces
that \tname{} uses to access the feature model to check legality of
configurations.  The type \CodeIn{FeatureVar} denotes a feature
variable.  A \CodeIn{VarAssign} object encodes an assignment of
boolean values to feature variables.  An assignment can be
\definit{complete}, assigning values to all the features, or
\definit{partial}, assigning values to a subset of the features.  A
complete assignment is \definit{valid} if it satisfies the constraints
of the feature model.  A partial assignment is \definit{satisfiable}
if it can be extended to a valid complete assignment.  

The \CodeIn{FeatureModel} interface provides queries for determining the
validity of feature assignments, obtaining valid configurations, and
checking if particular informed features are mandatory. Given an
assignment $\alpha$, the method \CodeIn{getValid()} returns the set of
all complete assignments that (1)~agree with $\alpha$ on the values of
feature variables in $\alpha$ and (2)~assign the values of the
remaining feature variables to make the complete assignment valid.  If
the set is not empty for $\alpha$, we say that $\alpha$ is
\definit{\satisfiable}; the method \CodeIn{isSatisfiable()} checks
this. The method \CodeIn{isMandatory()} checks if a feature is
mandatory according to the feature model and the method
\CodeIn{getMandatoryValue()} returns the mandatory value for the
informed feature.  We use the SAT4J~\citep{sat4j} SAT solver to
implement these feature model operations.

\subsubsection{Algorithm}

Figure~\ref{fig:splKoratModified} shows the pseudo-code of
\SPLKorat{}, modified to support evolution.  \SPLKorat{} takes on
input a feature model, a test, and an optional stack denoting what
decision subtree should be explored.  The modified statements appear
underlined.  Figure~\ref{fig:evoSPLKoratAlgorithm} shows the
pseudo-code of \tname{}, which calls \SPLKorat{}.
\RegressionSPLKorat{} takes as input a feature model, a test, and a
method list (line~\ref{definition}), empty on the first run, denoting
those methods that have been impacted by evolutionary changes.  The
first run of \tname{} invokes \SPLKorat{} passing an empty stack on
input.  In fact, this initial call is equivalent to invoking the
non-evolutionary version of \SPLKorat{}.  Subsequent runs of \tname{}
invoke \SPLKorat{} passing non-empty stacks corresponding to subtrees
of the exploration tree that need to be explored upon changes.  In the
following we elaborate how the \tname{} works on the first run and
subsequent runs.

\subsubsection*{First Run}
\label{sec:firstrun}

\SPLKorat{} maintains a \CodeIn{state} (line~\ref{state2}) that stores
the values of feature variables, and a \CodeIn{stack}
(line~\ref{stack2}) of \CodeIn{Entry} types (line~\ref{entry-types}),
representing methods and feature variables accessed during the test
run.  For each read of an optional feature variable, \SPLKorat{} calls
\CodeIn{notifyFeatureRead()} (line~\ref{notifyFeatureRead}) to update
both the \CodeIn{stack} with the new feature variable read and the
\CodeIn{state} with a satisfiable value.  For each method read, the
algorithm calls \CodeIn{notifyMethodVisited()}
(line~\ref{notifyMethodVisited}) to update the \CodeIn{stack} with the
new invoked method.  At line~\ref{runexecution1}, \SPLKorat{} executes
the test on a valid partial assignment.  After finishing execution,
\SPLKorat{} determines the next partial assignment to execute (lines
\ref{nextStart2}--\ref{nextEnd2}).  If the last read feature has value
\CodeIn{true}, then \SPLKorat{} has explored both values of that
feature, and it is popped off the stack
(lines~\ref{nextPop1}--\ref{nextPop2}).  If the size of the stack
becomes smaller than the size of the stack passed on input
\CodeIn{st}, it means the subtree of interest has been fully explored
and \SPLKorat{} should terminate.  If the last read feature has value
\CodeIn{false}, then \SPLKorat{} has explored only the \CodeIn{false}
value, and the feature should be set to \CodeIn{true}
(lines~\ref{nextTrue1}--\ref{nextTrue2}). Another important step
occurs now (line~\ref{nextValid}).  While the backtracking over the
\CodeIn{stack} found a partial assignment to explore, it can be the
case that this assignment is not \satisfiable{} for the feature
model. In that case, \SPLKorat{} keeps searching for the next
\satisfiable{} assignment to run.  If no such assignment is found, the
\CodeIn{stack} becomes empty, and \SPLKorat{} terminates.

\begin{figure}[t!]
  \input{splat-new}
  \caption{\label{fig:splKoratModified}\SPLKoratName{} algorithm modified.}
\end{figure}
\begin{figure}[t!]
  \input{evo-splat}
  \caption{\label{fig:evoSPLKoratAlgorithm}\RegressionSPLKorat{}
    algorithm (\rcs{}).}
\end{figure}

\vspace{1ex}\noindent\textbf{Caching SAT calls.}  Assuming that the subject under
test provides a feature model that constraints the set of legal
configurations, \tname{} reduces execution cost by caching the results
of SAT solver calls (see lines~\ref{nextValid} and \ref{notify1}).
This caching is the simplest memoization mechanism offered by
\tname{}; it focuses on time reduction.  To implement this mechanism
\tname{} uses a map, where keys correspond to partial variable
assignments and values correspond to booleans indicating whether or
not the corresponding assignment is satisfiable.  Given that these
keys are potentially highly similar by construction \tname{} uses a
trie (aka prefix trees)~\citep{Fredkin-Trie} for faster storage and
lookup of key-value pairs.  Given the fact that generated constraints
are of the form $\bigwedge{}a_i$, with $a_i=v$ or $a_i=\neg{}v$ for a
symbolic variable $v$, canonicalization is obtained by pre-determining
an ordering on symbolic variables.  For example, \tname{} will
represent the constraint $y\wedge{}x$ as $x\wedge{}y$, given the name
order $x<y$.



\subsubsection*{Next Runs}
\label{sec:nextrun}

The current version of \tname{} assumes that potentially
fault-revealing changes occur inside method bodies.  Under this
assumption it is able to apply very lightweight change-impact
analysis, which \tname{} builds on.  Considering that all tests passed
in the previous evolution cycle, a call to a changed method is a
necessary condition to activate potential failures.  As such, \tname{}
monitor calls to changed methods to decide what configurations needs
to be re-executed on each test.  

  

\RegressionSPLKorat{} explores configurations as follows.  It first
initializes the values of feature variables (lines
\ref{initStart1}--\ref{initEnd1}). The algorithm then instruments
(line~\ref{instrument1}) the code under test to observe both feature
variables read and methods calls.  Next, it reads the list of traces
(line~\ref{cache}) observed during the previous run of \tname{} for
that test.  If the list is empty, it means no history is available and
\SPLKorat{} is called for a full-run (line~\ref{first-splat-call}).
Otherwise, \RegressionSPLKorat{} calls \SPLKorat{} to explore all
impacted configurations (lines~\ref{foreach-begin}-\ref{foreach-end}).
Each iteration in this loop calls \SPLKorat{} on one decision subtree
that has been affected by changes.  The parameter \CodeIn{st} in the
call at line~\ref{splat-call} denotes such subtree; the object
\CodeIn{st} encapsulates a partial assignment of feature variables to
values.  For example, the assignment \CodeIn{[(WEIGHTED,1)]} denotes
the right subtree from the example in Section~\ref{sec:space-red}.




\subsection{\tname{} for \rts{}}
\label{sec:evosplat-rts}

In contrast to \rcs{}, \rts{} does not restrict configurations for
execution in each test.  Instead, \rts{} restricts the tests that will
be executed in each evolution cycle.  For those tests, all
reachable configurations should be executed.  Note that \rts{} is a
particular case of \rcs{}~--~one could model the effects of \rts{}
with \rcs{} by assigning an empty set of configurations to those tests
that should be ignored and retaining the original set of
configurations for selected (non-ignored) tests.  

Despite the generality of \rcs{} (\ie{}, \tname{} for \rcs{}), there
are scenarios where configuration selection has limited effectiveness.
\rts{} comes as an alternative for those cases.  More precisely,
reduction in number of configurations with \rcs{} is limited when most
configurations explored in tests cover most functions.  In those
cases, it is possible that the changed function(s), if covered by the
test at all, will likely be covered by any of its reachable
configurations.  Situations like this arise more frequently in systems
whose tests exercise a single main
function~\citep{medeiros-etal-icse2016}.  \gcc{} is a case in point
(see Section~\ref{sec:eval_rts}).


To support \rts{}, \tname{} first identifies what tests have been
impacted by changes and then performs a full execution with
\SPLKorat{} on those tests.  \tname{} maintains a map from tests to
set of functions that have been covered by any configuration reachable
from that test.  Recall that, in this scenario, most configurations of
a given test cover most functions.  At test selection time, a test
will be considered for testing if a changed function is included in
the set associated to that test.


\section{Evaluation}
\label{sec:eval}

To reflect our different scenarios of evaluation, this section is
structurally organized in two parts, as Table~\ref{eval:organization}
shows.

\setlength{\tabcolsep}{2.5pt}
\begin{table}[h!] 
  \footnotesize
  \centering
  \begin{tabular}{|c|c|c|c|c|} 
    \hline
    \multirow{2}{*}{Section} & \multirow{2}{*}{Scenario} & \multirow{2}{*}{RQs} & Subject & \multirow{2}{*}{Type of Tests} \\
    & & & Programs &  \\    
    \hline
    \hline
    \ref{sec:eval_rcs} & RCS & R1, R2 & SPLs & Unit/System/Integration \\
    \ref{sec:eval_rts}  & \rts{} & R3, R4, R5 & GCC & System\\
    \hline
  \end{tabular}
  \vspace{-2ex}
  \caption{\label{eval:organization}Organization of  Section~\ref{sec:eval}.}
\end{table}
\normalsize

Recall that in Regression Configuration Selection (\rcs{}) mode
\tname{} aims to reduce the number of configurations reachable by each
test (but with the test set fixed) whereas in Regression Test
Selection (\rts{}) mode \tname{} aims to reduce the number of tests
executed (but with configurations of tests fixed).

Section~\ref{sec:limitation} reports on a limitation study we
conducted to decide when to apply \rcs{} versus \rts{}.
Section~\ref{sec:spls} evaluates \tname{} under the scenario of \rcs{}
(see Section~\ref{sec:evosplat-rct}) on twelve publicly-available
\Comment{relatively-small} Software Product Lines (SPLs) written in Java.
Section~\ref{sec:gcc} evaluates \tname{} under the scenario of \rts{}
(see Section~\ref{sec:evosplat-rts}) on \gcc{}, a large configurable
system built over the course of 3 decades.\Comment{ This section
  focuses on RQ3.}

\subsection{Limitation Study: \rcs{} vs. \rts{}}
\label{sec:regression}\label{sec:limitation}

Although both SPLs and configurable systems use variability as key
principle, the characteristics of the tests in these programs are
different.  As observed in recent empirical
studies~\cite{abal-etal-ase2014,medeiros-etal-icse2016}, most tests in
configurable systems are system tests, which often cover large
portions of the code.  This limits the ability of \rcs{} to prune
configurations.  The intuition is that \tname{} for \rcs{} would
benefit the most in a scenario where, for a given test, distinct
configurations cover highly-different sets of functions.  In that
scenario, if a function changes, only few configurations would be
impacted and only few re-executions would be necessary for a given
test.  If, on the contrary, most functions were covered across most
configurations, \rcs{} would have little effect as the test would need
to be re-executed in all those configurations. In this later scenario
\tname{} for \rts{} is preferable. It is unable to prune configuration
but is more lightweight.





\begin{figure}[t!]
\centering
\vspace{-5ex}
\includegraphics[width=.4\textwidth]{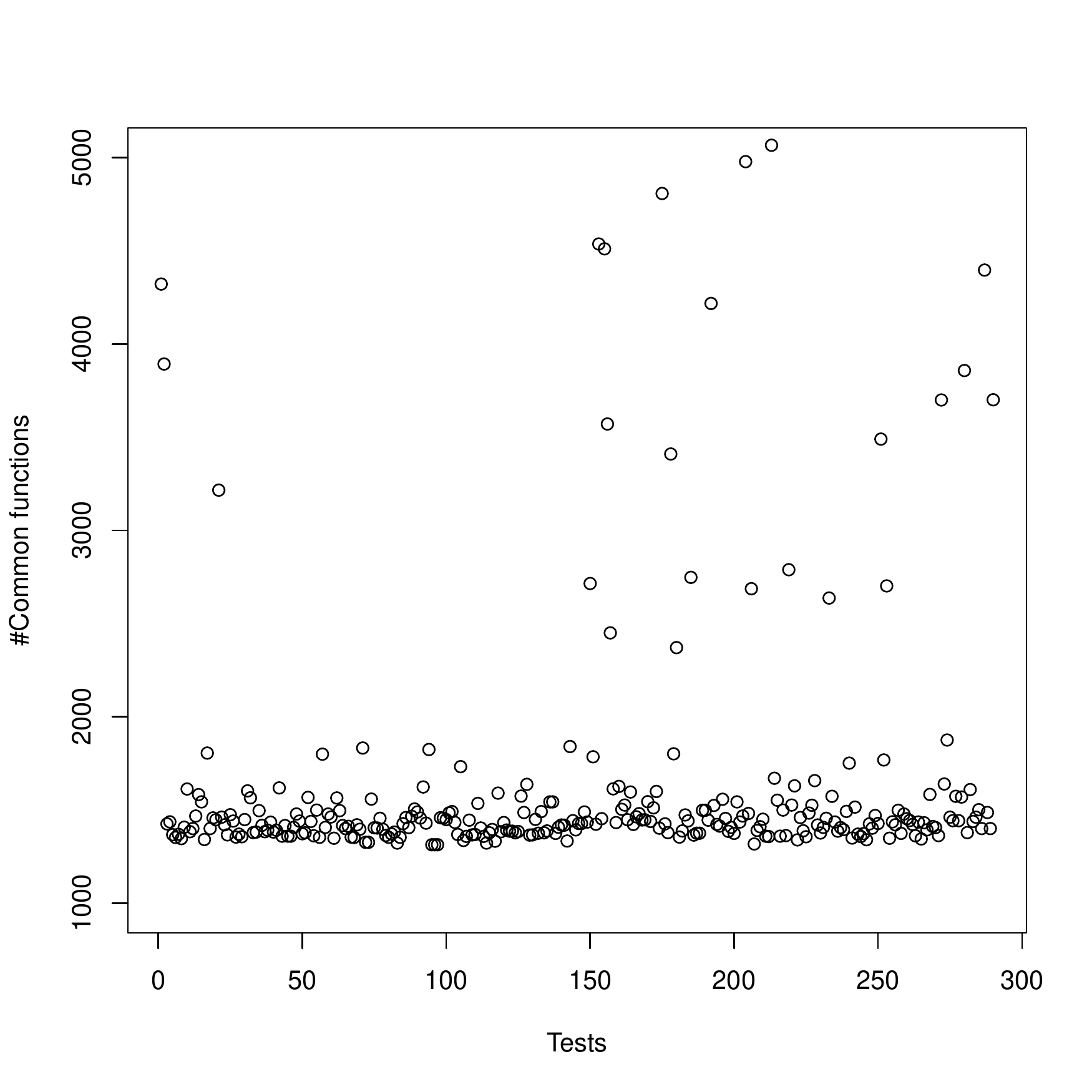}
\vspace{-3.5ex}
\caption{Distribution of common functions across configurations for
  each test.}
\label{fig:pfap}
\end{figure}


We conducted a limitation study to understand how successful \rcs{}
would be in a configurable system like \gcc{}.  More precisely, we
studied the variance in the distribution of functions covered across
configurations of each test in \gcc{}.  Figure~\ref{fig:pfap}
illustrates the distribution of functions that are common across all
reachable configurations.  The average amount of functions per
configuration lies between one and two thousand\Comment{, except the
  outliers}.  A closer inspection reveals that most configurations of
a given test cover about the same set of functions.  This happens
because the test in \gcc{} exercise a main function, which dynamically
reaches most functions of the compiler (see Section~\ref{sec:deja}).
This is common practice in configurable systems where configuration
options are passed on the command line.  We also observed a high
percentage of common functions across configurations and across tests,
72\% of functions are called by all tests.

Given these observations, we evaluated \tname{} for \rts{} with \gcc{}
and evaluated \tname{} for \rcs{} with \spl{}s.



\subsection{Regression Configuration Selection (\rcs{})}
\label{sec:spls}\label{sec:eval_rcs}

Considering the scenario of \rcs{}, evaluated against Software Product
Lines (\spl{}s), we posed the following research questions:

\begin{itemize}
\item[\textbf{RQ1}.]~What are the time savings obtained with the
  caching of constraint solver calls?
  
\item[\textbf{RQ2}.]~What are the space savings obtained with the
  recording of decision trees?
\end{itemize}




For this experiment we selected \numsubjects{} subjects of various
sizes that have been previously used in other
studies. Table~\ref{tab:subjects} provides details about these
subjects, including number of features, number of valid configurations
that can be generated combining these features, number of methods, and
code size.

\setlength{\tabcolsep}{2pt}
\renewcommand{\arraystretch}{0.9}
\begin{table}[h!] 
\footnotesize
\centering
\begin{tabular}{cccccc}
\toprule
\emph{Subject} & \emph{Source} & \#\emph{Features} & \#\emph{Valid Confs.} & \#\emph{Methods} & \emph{LOC}  \\
\midrule
101Companies    & \citep{companies}                 &  10  &  192 & \z307 &   \z2,059  \\ 
DesktopSearcher & \citep{Philipp-vamos14}           &  16  &  462 & \z113 &   \z3,779  \\ 
Elevator        & \citep{elevator}                  & \z5  & \z20 & \z121 &   \z1,046  \\ 
Email           & \citep{email}                     & \z8  & \z40 & \z227 &   \z1,233  \\ 
GPL             & \citep{Lopez-herrejon01astandard} &  13  & \z73 & \z\z95&   \z1,713   \\
JTopas          & \citep{jtopas}                    & \z5  & \z32 & \z578 &   \z2,031  \\ 
MinePump        & \citep{mine}                      & \z6  & \z64 & \z\z99 & \z\z580  \\ 
Notepad         & \citep{chpkim-rv2010}             &  17  &  256 & \z136 &   \z2,074  \\
Prevayler       & \citep{prevayler}                 & \z5  & \z32 & \z523 &   \z2,844  \\
Sudoku          & \citep{sudoku}                    & \z6  & \z20 & \z\z99 & \z\z853  \\ 
XStream         & \citep{xstream}                   & \z7  &  128 & 2,318 &    14,480  \\
ZipMe           & \citep{apel-beyer-icse2011}       &  13
& \z24& \z338 &   \z3,650  \\
\bottomrule
\end{tabular}
\vspace{-1ex}
\caption{\label{tab:subjects}SPLs used.}
\end{table}
\normalsize

We classify execution of \tname{} in one of three categories:

\begin{itemize}
\item \textbf{Complete Re-Execution.}  All tests and configurations
  have to be re-executed.  This is necessary to bootstrap \tname{},
  for example. (Section~\ref{sec:complete-reexec}.)
\item \textbf{Partial Re-Execution.} Some tests and configurations
  have to be re-executed because only some of them were affected by
  changes.  (Section~\ref{sec:partial-reexec}.)
\item \textbf{No Re-Execution.} No tests (and configurations) need to
  re-run because none of them were affected by changes.
\end{itemize}

Our evaluation on \spl{}s considers the first two scenarios of
regression testing.  We do not discuss the third scenario as there is
no cost associated with running \tname{}.  We use the first scenario
to answer RQ1 and the second scenario to answer RQ2.

\subsubsection{Complete Re-execution (Time Reduction)}
\label{sec:complete-reexec}

A complete re-execution may be necessary during evolution.  Consider,
for example, the first run of the test suite using \tname{} or the
scenario where all tests and configurations have been impacted by
changes.  This experiment evaluates the conjecture that \tname{} can
help even in such extreme cases.



\tname{} optimizes the cost of SAT solving.  Recall that \SPLKorat{}
can use a SAT solver to discard invalid configurations.  On the one
hand, this is important because often only a small fraction of the
$2^{\#\mathit{Features}}$ configurations are valid.  See columns
``\#\emph{Features}'' and ``\#\emph{Valid Confs.}'' in
Table~\ref{tab:subjects}.  This indicates that using a SAT solver
conceptually pays off.  On the other hand, the cost of SAT solving can
be substantial.  This indicates that caching results of SAT solver
calls may be important. 


This experiment evaluates the distinct benefit of caching the calls to
the SAT solver~(lines~\ref{satcache-starts}--\ref{satcache-ends} from
Figure~\ref{fig:splKoratModified}).  Recall that \tname{} uses a trie,
shared across all tests, to store the outcomes of feasibility checks
and that trie keys correspond to feature constraints.  To reduce noise
in measurements, we discarded subjects from Table~\ref{tab:subjects}
whose test suite takes less than five seconds to run.

To isolate cost of SAT solving, this experiment simulates the scenario
where code changes would not influence feature decisions.  Under these
circumstances the decision trees generated in consecutive runs of
\tname{} on the same test would be the same.  This means that only
one, potentially expensive, \SPLKorat{} run is necessary until the
point where a change in the decision tree is observed.  Until that
point it suffices to execute a test against each of the configurations
obtained with the first execution of \SPLKorat{} on that test.  In
this setup, it is possible to isolate the benefits of an optimal cache
hit ratio.  Although we did not measure how prevalent this scenario is
to the practice, we conjecture that it happens often.  If, on the
contrary, decision trees were to always change during evolution, that
would certainly mean a lack of understanding about the purpose of
options.  

We considered three techniques for comparison:
\vspace{-1.3ex}
\begin{itemize}
\item ``\SPLKorat{}'' corresponds to the baseline technique.
\item ``FirstRun'' corresponds to the technique that runs \tname{} for
  the first time, with an empty cache, as described in
  Section~\ref{sec:firstrun}.
\item ``NextRuns'' corresponds to the technique that runs \tname{}
  with a filled cache, as described in Section~\ref{sec:nextrun}, but
  assuming the decision tree does not change.
\end{itemize}
\vspace{-1ex}

\setlength{\tabcolsep}{2pt}
\begin{table}[h!] 
\footnotesize
\centering
\begin{tabular}{crrrrr}
\toprule
\multirow{2}{*}{\emph{Subject}} & \multicolumn{3}{c|}{\SPLKorat} & \multicolumn{2}{c|}{\tname{}(\%)} \\
\cmidrule(r){2-6}
& \#SAT & \#UNSAT & TotalTime~(CS\%) & FirstRun & NextRuns \\
\midrule
101Companies    & 10,187 & 642 & 933 (0.60) & 0.27 & 0.31 \\ 
DesktopSearcher & 4,942 & 272 & 1,261 (14.80) & 5.80 & 14.35 \\ 
Email           & 5,446 & 432 & 6.7 (73.00) & 19.40 & 46.27 \\ 
GPL             & 4,652 & 420 & 7.18 (89.00) & 5.30 & 53.14 \\ 
JTopas          & 788 & 0 & 139 (9.85) & 3.70 & 5.17 \\ 
Notepad         & 168,549 & 18,695 & 16,025.5 (30.00) & 6.30 & 22.50 \\ 
XStream         & 486 & 0 & 93.8 (15.00) & 10.80 & 12.00\\ 
\midrule
\textbf{AVG}    & - & - & - (33.17) & 7.36 & 21.96 \\
\bottomrule
\end{tabular}
\vspace{-1ex}
\caption{\label{tab:savings-constraintsolving}Savings in SAT calls.}
\vspace{-2.5ex}
\end{table}
\normalsize

Table~\ref{tab:savings-constraintsolving} summarizes results for this
experiment.  Recall that \tname{} stores constraints in a trie and
that, to maximize cache hit ratio, the keys in the trie are stored in
a canonical form (Section~\ref{sec:firstrun}).  This figure shows the
breakdown of SAT queries that the baseline
technique~--~\SPLKorat{}~--~generates and land in SAT or UNSAT.  Note
that the number of SAT calls can be very high in some cases.  That
happens because (Evo)\SPLKorat{} checks consistency of partial
configurations incrementally. Exploration backtracks only when the
solver answers UNSAT to a query, which means it is no longer possible
to reach a valid configuration.  The other solver calls will result in
a SAT answer.


  
Column ``TotalTime~(CS\%)'' shows the overall time in seconds.  This
comprehends time spent running all tests on all (valid)
configurations, and the time spent with constraint solving.  The
percentage of time spent with constraint solving alone ``(CS\%)''
appears in parentheses.  Column ``FirstRun'' reports the reduction in
time obtained with the initial execution of \tname{}.  Saving of time
happens even in this case because \tname{} starts using the trie as it
is filled.  Column ``NextRuns'' reports the reduction in time obtained
with the subsequent executions of \tname{}.  Recall that in this case
\tname{} makes no calls to the SAT solver.  Note from the table the
difference between the savings obtained and the maximum possible
savings (column ``\%CS'').  For instance, considering XStream, the
savings of both ``FirstRun'' and ``NextRuns'', albeit significant
(10\% and 12\%, respectively), are still below the maximum possible
(15\%).


Considering all subjects, tests, and configurations that we analyzed,
results show an average time reduction of 21.96\%.  Results also show
that this value represents 66.2\% (21.96/33.17) of the maximum
possible reduction, which is the total time spent with constraint
solving in \SPLKorat{}.  It should be possible to reduce this gap
further by using more efficient canonicalizations and trie
implementations.

Time savings seem to depend on several factors such as complexity of
constraints, number of paths, and length and complexity of tests.  As
expected, we noted smaller improvements for the cases where relatively
more time is spent executing test code as opposed to calling the SAT
solver.  For example, the tests for the subjects 101Companies,
DesktopSearcher, and Notepad, spend a considerable time accessing the
graphical user interface and the tests for the subjects JTopas and
XStream spend a significant amount of time accessing files.

\begin{center}
  \fbox{
    \begin{minipage}{8cm}
      \textit{Answering RQ1:}~Results indicate that caching results of
      SAT solver calls should be done, although performance of caching
      depends on the characteristics of the application.
      \end{minipage}
    }
\end{center}




\subsubsection{Partial Re-execution (Space Reduction)}
\label{sec:partial-reexec}
\label{exp:changes-app-code}

A partial re-execution is necessary when changes affect only some
configurations reachable from given tests. In this case, \tname{}
reconstructs only the parts of the decision tree that may have been
affected by changes. This experiment evaluates how much \tname{} can
reduce space required to explore configurations. More specifically, we
want to find out how much less configurations, compared to a complete
run (as in Section~\ref{sec:complete-reexec}), one would need to
re-run in a scenario of change.



This experiment assumes that changes occur only in methods and, within
method bodies, changes occur in application code, not in feature
decisions, and that feature variables are not re-assigned within code
(as in the SPLs we analyzed).  Under these circumstances the decision
tree, obtained in previous runs, is not affected.  Intuitively, this
setup enables one to filter subtrees~--~inducing a set of
configurations~--~affected by changes.  Furthermore, to limit the
scope of the experiment,
we are assuming changes in only one method per
evolution cycle.  The rationale for that is that the size of the SPL
subjects we analyzed is small (see Table~\ref{tab:subjects}). Note
that, albeit small, these are subjects typically used to study SPL
testing. Section~\ref{sec:gcc} reports results of \tname{} on a much
larger software system.

\setlength{\tabcolsep}{1pt}
\begin{table}[t!]
\scriptsize
\centering
\begin{tabular}{ccccccccc}
\toprule
\emph{Subject} & [0-0.1[ & [0.1-0.2[ & [0.2-0.3[ & [0.3-0.4[ & [0.4-0.5[ & [0.5-0.6[ & [0.6-0.7[ & \textbf{Total}\\ 
\midrule
101Companies       & \z9 & \z2 & \z4 & \z4 & \z3 & - & - & 22 \\ 
DesktopSearcher    & \z1 & - & \z5 & \z8 & \z30 & - & - & 44 \\ 
Elevator           & \z1 & \z2 & - & - & - & - & - & \z3 \\ 
Email              & \z1 & - & \z6 & \z7 & \z1 & - & - & 15 \\ 
GPL                & \z8 & \z4 & \z1 & \z1 & \z7 & \z3 & \z1 & 25 \\ 
JTopas             & 28 & - & - & - & - & - & - & 28 \\ 
MinePump           & - & - & \z2 & \z1 & - & - & - & \z3 \\ 
Notepad            & - & - & 42 & \z1 & - & - & - & 43 \\ 
Prevayler          & - & \z2 & \z3 & \z1 & \z2 & - & - & \z8 \\ 
Sudoku             & - & \z1 & - & - & \z4 & - & \z1 & \z6 \\ 
XStream            & \z7 & - & - & - & - & - & - & \z7 \\ 
ZipMe              & 14 & 18 & \z7 & \z3 & \z5 & - & - & 47 \\
\midrule
\textbf{Total}     & 69 & 29 & 70 & 26 & 52 & 3 & 2 & - \\ 
\bottomrule
\end{tabular}
\caption{\label{tab:distributions} Distribution of tests per 
  subject over increasing intervals of configuration reduction.  A
  configuration reduction for a given test is a measure of how
  much \tname{} reduces the number of configurations explored (and
  hence cost of exploration) compared to running the test again
  with \SPLKorat{}.  For example, of the 25 tests of GPL, 7 tests lie
  in the [0.4-0.5[ interval.  This means that the number of
  configurations \tname{} explores would be reduced by a factor of
  $\sim$45\% for those tests.  The rightmost interval clusters the
  test cases that are most beneficial for \tname{} in number of
  configuration savings whereas the leftmost interval clusters the
  worst-performing cases.  Note that this setup ignores changes that
  could affect the decision tree (and produce less/more traces).}
\end{table}
\normalsize
\setlength{\tabcolsep}{6pt}

We used the following procedure to measure configuration reduction.
We run \SPLKorat{} once on each test and measure, for each method that
appears in the trace, the fraction of configurations whose test
execution trace does not contain that method.  The intuition is that
it is not necessary to run the test on a configuration that does
\emph{not} call the changed method.  For a given test case, we save
the average saving across all methods.  For the sake of illustration,
let us assume that a test explores only two configurations, $c_1$ and
$c_2$, that the code contains only three methods $m_1$, $m_2$, and
$m_3$, and that configuration $c_1$ covers $m_1$ and $m_2$ whereas
configuration $c_2$ covers $m_1$ and $m_3$.  For this setup a change
(only) in method $m_1$ results in no reduction as it is always
referred in the traces while a change in either $m_2$ or $m_3$ results
in 50\% reduction (in both cases a single configuration is
eliminated).  An average reduction of 33\% will be observed
considering the three methods.

Table~\ref{tab:distributions} presents the distribution of tests per
subject over different intervals of configuration reduction.  The
higher the interval a test is allocated the better.  We omitted
columns where no test cases felt under. For example, considering this
setup, \tname{} reduces the number of configurations explored by an
average of $\sim$45\% (cf. interval ``[0.4,0.5['') for 7 tests of GPL.

\begin{center}
  \fbox{
    \begin{minipage}{8cm}
      \textit{Answering RQ2:}~Considering all subjects in this
      experimental setup, results indicate that it is possible to
      obtain a significant reduction in number of configurations
      explored.  However, this highly depends on the functions changes
      and the impact set associated to those functions.
      \end{minipage}
    }
\end{center}




\subsection{Regression Test Selection (\rts{})}
\label{sec:gcc}\label{sec:eval_rts}

This section describes a case study involving the GNU Compiler
Collection (\gcc)~\citep{gcc}, a large system with more than 7 million
lines of code, more than 17k tests, with hundreds of configuration
options~\citep{gccoptions}, and 2,015 feature variables.  GCC has been
developed for almost three decades, receiving contribution from over
500 developers~\citep{gccdoc}.  

The goal of this study is to assess how \tname{} performs on a
scenario of \rts{} using a highly-configurable real system.  We pose
the following research questions:

\begin{itemize}
\item[\textbf{RQ3}.]~How efficient \tname{} is compared to the \SPLKorat?
\item[\textbf{RQ4}.]~How efficient \tname{} is compared to sampling
  techniques?
\item[\textbf{RQ5}.]~How effective \tname{} is (w.r.t. bug finding)
  compared to alternative techniques?
\end{itemize}






\subsubsection{Techniques, Metrics, and Methodology}
\label{sec:methodology}

\textbf{Techniques.}~We evaluated \tname{} against \SPLKorat{} (the
baseline technique) and sampling techniques, namely
$\mathit{t}$-$\mathit{wise}$ sampling for $t$=2 and $t$=6.  These
values of $t$ have been recently used by Medeiros
\etal{}~\cite{medeiros-etal-icse2016} to evaluate low and high bounds
of $\mathit{t}$-$\mathit{wise}$ testing in configurable systems.


\textbf{Metrics.}~The metrics we used to evaluate efficiency of
techniques are \emph{number of tests} selected for re-execution upon
an evolution cycle and \emph{time reduction} The metric used to
evaluate efficacy is \emph{fault-detection}.  For that, we considered
five bugs we previously found in \gcc{} version
\gccversion{}~\cite{sabrina-etal-splc15}.

\textbf{Methodology.}~To bootstrap \rts{} one needs to first map the
functions covered by each test on any reachable configuration.  In a
subsequent iteration, if any of the functions covered by a given test
changes, \tname{} for \rts{} (\tname{} for short) re-runs that test
against \emph{all its configurations}.  Otherwise, \tname{} skips the
test.  For example, \SPLKorat{} reaches 192 configurations for the
test \CodeIn{gcc-dg.pr58145-2.c}.  Executing this test against these
configurations cover 4,129 functions. Analyzing the evolution cycle
the day after August 21, 2015 (our day of reference), we observed that
the changes impacted all the 192 configurations.  More precisely, for
any given reachable configuration of the test, at least one of the 31
modified functions in this evolution cycle appeared in its
corresponding execution trace.  As result, \tname{} re-executed the
test on all 192 configurations.

\subsubsection{Infrastructure}
\label{sec:deja}





\gcc{} uses
DejaGnu~\citep{dejagnu} as a testing framework and runs each test
against a single configuration.  Figure~\ref{fig:test_gcc} shows an
example of \gcc{} test, which is comprised of the following parts.

\definecolor{mygray}{rgb}{0.5,0.5,0.5}
\definecolor{mymauve}{rgb}{0.58,0,0.82}
\begin{figure}
  \begin{subfigure}
  \centering
  \scriptsize
  \lstset{language=C++,language=C, frame=single, basicstyle=\ttfamily\scriptsize, keywordstyle=\color{blue}, stringstyle=\color{mymauve}, commentstyle=\itshape\color{mygray}, boxpos=c, numbers=left, xleftmargin=5.0ex, framexrightmargin=0pt, framexleftmargin=5pt, framexbottommargin=1pt, numberstyle=\tiny}
    \begin{lstlisting}
/* { dg-do compile } */
/* { dg-options "-O2 -fdump-tree-optimized" } */
int g(_Complex int*);
int f(void){
  _Complex int t = 0;
  int i, j;
 __real__ t += 2;
 __imag__ t += 2;
  return g(&t);
}
/* { dg-final { scan-tree-dump-times "__complex__" 0 "optimized" { xfail *-*-* } } } */
    \end{lstlisting}
    \vspace{-2ex}
    \caption{\label{fig:test_gcc}An example of test (complex-4.c from GCC test suite).}
  \end{subfigure}
  \par\vfill
  \vspace{3ex}
  \begin{subfigure}
    \centering
	 \includegraphics[width=.48\textwidth]{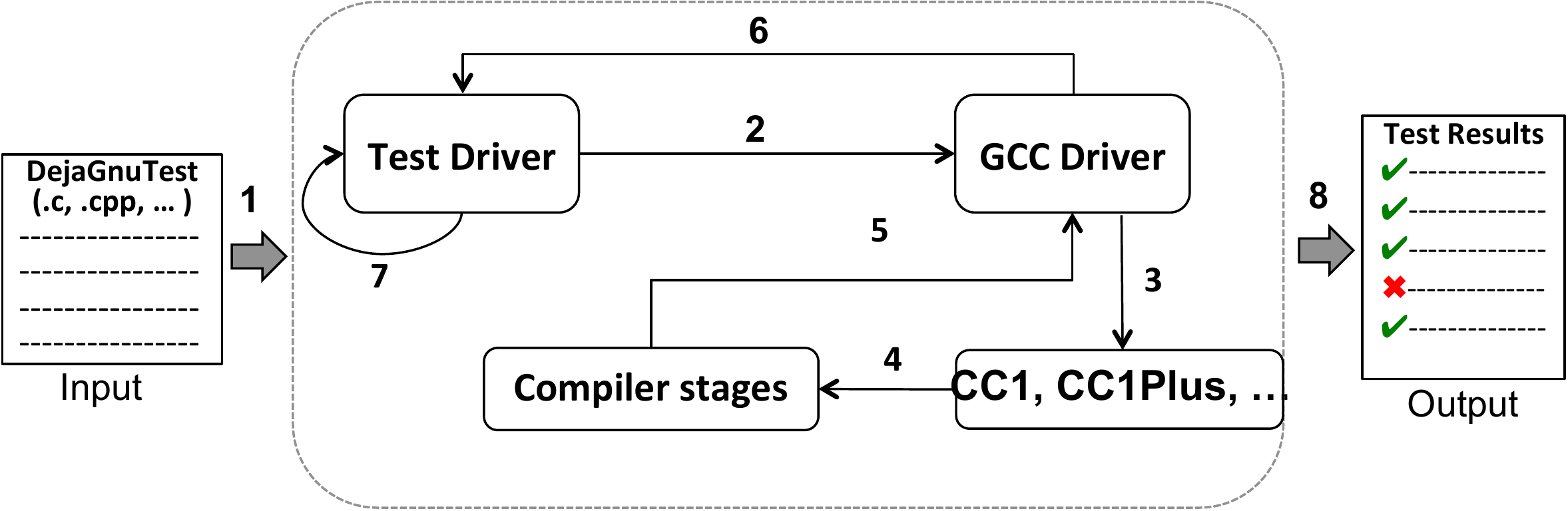}
    \vspace{-4ex}
    \caption{\gcc{} test execution workflow.}
    \label{fig:test-process}
  \end{subfigure}
\end{figure}

\begin{itemize}

\item \emph{Actions}: The DejaGnu directive (\emph{dg-do}) specifies
  the type of the test. For \gcc{}, examples of actions are:
  preprocess, compile, assemble, link, or run. For example, the
  directive \CodeIn{dg-do compile} (Line~1) instructs DejaGnu to only
  compile functions \CodeIn{g} and \CodeIn{f}.  No other
  operations (\eg{}, assemble, link, run, etc.) will be executed.

\item \emph{Options}: The Dejagnu directive (\emph{dg-options})
  specifies a subset of options that must be present to run the test.
  For this example, the directive at line 2 indicates that this test
  must be executed with at least the options "O2" and
  "-fdump-tree-optimized" enabled.

\item \emph{Code}: For the case of \gcc{} the body of the test
  corresponds to code.  More precisely, C/C++ code. In this example,
  the C code appears from lines 3 to 10;

\item \emph{Outcomes}: The instruction that appears at line 11
  determines whether the test passes or fails.
\end{itemize}


Figure~\ref{fig:test-process} illustrates the test execution workflow
for GCC.  The input of the workflow is a test (as in
Figure~\ref{fig:test_gcc}) and the output is the report of test
results. After reading the test (1), the \emph{Test Driver} calls the
\emph{GCC Driver} passing the commands interpreted from the test file
(2).  Based on the test input, the \emph{GCC Driver} calls the
\CodeIn{main function} of the compiler (\eg{}, CC1 for C, CC1Plus for
C++) passing compilation parameters (3).  Different compiler stages
(scan, parse, etc.)  will be triggered depending on the ``Actions'' of
the test (4).  The compiler sends results back to the \emph{GCC
  Driver} when compilation finishes (5) and the \emph{GCC Driver}
forwards the output to the \emph{Test Driver} (6), which checks these
outputs (7) and elaborates a test report (8).

Other configurable systems are tested similarly to \gcc{}, using the
main function for system testing~\citep{medeiros-etal-icse2016}. This
is important as to decide which reduction strategy to use in
regression testing (see Section~\ref{sec:regression}).

\subsubsection{Experimental Setup}
\label{gcc:setup}


\textbf{Tests.} Test execution in \gcc{} is
time-consuming. Considering one configuration per test and our running
environment, it takes roughly 2s to run each test. In this experiment
we focused on the \CodeIn{gcc-dg} test suite, which contains 3,343
tests.  The rationale for choosing this test suite was that we
observed from bug reports a higher incidence of bugs revealed with
this suite.  


\textbf{Versions and Changes.}  Note that, in contrast to the
simulated scenario of evolution in the experiment involving software
product lines (Section~\ref{sec:spls}), here the scenario is real.  We
considered daily changes of \gcc{} as it represents an important
scenario of evolution.  We analyzed the build files that the \gcc{}
team posts to their web site every day.  Due to a relatively high cost
to prepare the execution environment (\eg{}, download, instrument
code, and build), we considered only seven evolution cycles, the
period of a week.  Figure~\ref{tab:changes} shows the amount of
functions that change over that period, starting on August 21, 2015.
We use the notation ``+$n$'' to indicate an evolution cycle from day
$n$ to day $n+1$.  For example, ``+1'' denotes the period between the
starting date and next day.  Note that the amount of changed functions
is very small relative to the total number of functions in \gcc{} (as
per Figure~\ref{fig:pfap}).  The stability of \gcc{} justifies that.


\textbf{Options.} To further reduce exploration time, we restricted
the number of configuration options that \splat{} considers
(see~\citep{gccoptions}). We used the 33 most-frequently cited options
from the \gcc{} bug reports in the week we focused our analysis.
 
\newcommand{\fsched}{\CodeIn{-fsched-spec-load}}
\newcommand{\fschedDangerous}{\CodeIn{-fsched-spec-load-dangerous}}
\newcommand{\fselschedPipelining}{\CodeIn{-fsel-sched-pipelining}}
\newcommand{\fselectiveSched}{\CodeIn{\nobreakdashes-fselective-scheduling}}
\newcommand{\fselectiveSchedTwo}{\CodeIn{-fselective-scheduling2}}
\newcommand{\flto}{\CodeIn{-flto}}
\newcommand{\fnoFatLtoObjects}{\CodeIn{-fno-fat-lto-objects}}
\newcommand{\fschedInsn}{\CodeIn{-fschedule-insns}}
\newcommand{\totalSabrinaConstraints}{136}
\newcommand{\totalGarvinConstraints}{110}
\newcommand{\totalManualConstraints}{246}

\setlength{\tabcolsep}{4.5pt}
\begin{table*}[t!]
  \small
  \centering
  \begin{tabular}{ccccccc}
  \toprule
    \emph{Evolution} & \emph{Functions Changed} & \emph{Tests Exec.~(\%)} & \multicolumn{2}{c|}{\tname{}} & \multicolumn{1}{c|}{\evopairwise{}} & \multicolumn{1}{c}{\evosixwise{}}\\
\cmidrule{4-7}
    &  &  & Confs/Test & Time~(h) & Time~(h) & Time~(h) \\ 
    +1 & 31  & 100   & 134   & 6.5 & 0.67 & 27.0  \\ 
    +2 & \z4 & 100   & 134   & 6.5 & 0.67 & 27.0  \\ 
    +3 & \z4 & \z\z- & \z\z- & \z\z- & \z\z- & \z\z- \\ 
    +4 & 15  & \z\z5 & 248   & 0.7 & 0.04 & \z1.4 \\ 
    +5 & 13  & \z\z- & \z\z- & \z\z- & \z\z- & \z\z- \\ 
    +6 & \z1 & \z57  & 169   & 4.8 & 0.38 & 15.6  \\ 
    +7 & \z8 & 100   & 134   & 6.5 & 0.67 & 27.0  \\ 
\midrule
\textbf{AVG} & 10.85 & 51.7 & 117 & 3.6 & 0.35 & 14.0 \\
\bottomrule
  \end{tabular}
  \vspace{-2ex}
  \caption{\label{tab:basic-rts}\label{tab:changes}RTS Results.
  \emph{\#Tests Exec.~(\%)}: the percentage of tests executed; 
  \emph{\#Confs/Test}: the average on the number of configurations executed per test;
  \emph{Time~(h)}: the time spent to run the tests and configurations.
  \evopairwise{} generated 13 Confs/Test; \evosixwise{} generated 553
  Confs/Test.}
\end{table*}
\normalsize

\textbf{Initial Feature Model and Ground Truth.}  Recall that \splat{}
is able to constrain the set of configurations to explore when a 
formally-specified feature model exists.  Unfortunately, \gcc{} does
\emph{not} provide such model.  In this study we enriched the feature
model constructed by Garvin~\etal{}~\citep{garvin-etal-2013} that
documents \totalGarvinConstraints{} constraints from \gcc{}.  We
augmented the model with constraints that we manually extracted from
the online documentation~\citep{gcc-otmization-options}.  We found
\totalSabrinaConstraints{} new constraints not included in
Garvin~\etal{}'s model, resulting in a total of 246 constraints of
which 86\% relate to the 33 options we analyzed.

\textbf{SPLat version.} In this study, we used a custom version of
SPLat that communicates with the tested subject (\gcc{}, in this case)
through the file system (see Figure~\ref{fig:test-process}).  To
discover new configurations for execution, this version of SPLat reads
accessed variables from the execution traces generated by an
instrumented version of the subject program.  Details of the
instrumentation, interfaces used, and code can be found
elsewhere~\citep{sabrina-etal-splc15}.





\subsubsection{Efficiency Results: Comparison with Baseline}

\textbf{Configuration Bounds.} Considering a single configuration per
test, it takes $\sim$1h18m to run only the \CodeIn{gcc-dg} test suite.
To reduce overall time for running our experiments, we specified an
upper bound of 100 configurations to explore per test.  Of the 3,343
tests from \CodeIn{gcc-dg}, 290 tests do not reach this bound.  We
considered only those tests as to enable \tname{} to explore the
entire decision tree on a re-run.  Without this scheme, it would be
possible to bias the search by narrowing the configuration space in
the tree.  Note that the purpose of this restriction is to facilitate
analysis of results; \tname{} does support cases that reach
configuration bounds.


\setlength{\tabcolsep}{2.5pt}
\begin{table}[h!] 
  \small
  \centering
  \begin{tabular}{cccc}
    \toprule
    \multirow{2}{*}{\emph{Daily Build}} & \emph{Baseline} & \multicolumn{2}{|c|}{\tname{}} \\
    \cmidrule{2-4}
    & Time(h) & Tests Exec.(\%) & Time(h) \\
    +1 & 10.15 & 100 & 10.15 \\ 
    +2 & 10.15 & 100 & 10.15 \\ 
    +3 & 10.15 & - & - \\ 
    +4 & 10.15 & 7.6 & 0.78 \\ 
    +5 & 10.15 & 100 & 10.15 \\ 
    +6 & 10.15 & 47 & 5.1 \\ 
    +7 & 10.15 & 100 & 10.15 \\ 
    \midrule
    AVG & 10.15 & 65 & 6.64 \\
    \bottomrule
  \end{tabular}
  \vspace{-2ex}
  \caption{\label{tab:splat-rts}\splat{} vs. \tname{}.}
  \vspace{-1ex}
\end{table}

Figure~~\ref{tab:splat-rts} show results comparing \tname{} with our
baseline, \SPLKorat{}.  Results show that \tname{} is either very
beneficial or not at all.  In 4 out of 7 cases \tname{} obtained no
reduction.  For the rest of the cases, a significant reduction was
obtained.  On average, results indicate a reduction of $\sim$35\% in
the number of tests required for re-execution and of running time.  
The cases were all tests required re-execution (see 100\% under column 
``\emph{Tests Exec. \%}'') reveal that during the week under consideration, 
a ``basic'' function was changed.  This function is shared across all 
tests and configurations. For example, the function \CodeIn{compile\_file} 
from the file \CodeIn{toplev.c} was modified in the first evolution 
cycle and that function is called by every test for every configuration.

\begin{center}
  \fbox{
    \begin{minipage}{8cm}
      \textit{Answering RQ3:}~Results obtained in a scenario of
      evolution (a week of evolution in \gcc{} release \gccversion{})
      shows that running \tname{} (as opposed to rerunning \SPLKorat{}
      from scratch) saves time.  The average reduction of time
      considering this scenario was nearly 35\%.
      \end{minipage}
    }
\end{center}

\subsubsection{Efficiency Results: Comparison with Sampling}

\textbf{Test Bounds.} In contrast to the previous experiment, this
experiment does not restrict the number of configurations that
\tname{} explores.  The rationale is that limiting the number of
configurations would obviously favor \tname{} compared to high values
of $t$ (with $t$ denoting the width of configuration vectors in
$t$-$\mathit{wise}$ testing) and disfavor \tname{} compared to low
values of $t$.  For that, \SPLKorat{} is not restricted to
configuration bounds in this experiment.  However, to reduce
experimentation time to a reasonable figure, we limited the number
\emph{of tests} to 100.  We randomly selected those from the 3,343
tests of \CodeIn{gcc-dg}.

Figure~\ref{tab:basic-rts} reports efficiency results. Column
\emph{Evolution} shows the evolution cycle id. An evolution cycle
refers to the changes between two consecutive daily builds. Column
\emph{Functions Changed} shows the number of functions that changed in
a cycle. We omitted results for the baseline as it is virtually
constant across different cycles: it takes nearly 6.5h to execute all
tests against all reachable configurations. The columns \tname{},
pairwise, and six-wise show results for each of the techniques. Column
\emph{Tests Exec.(\%)} shows the percentage of tests in the test suite
that required execution, it is the same for all techniques. A higher
number in this column indicates low reduction. Columns ``Time~(h)''
show the runtime cost of running each of the techniques. For \tname{},
column ``Confs/Test'' shows the \emph{average} number of
configurations explored per test. For pair-wise and six-wise, which
are black-box approaches, this value is constant.  Pair-wise explores
13 configurations per test and six-wise explores 553 configurations
per test.  Note that when there are code changes which are not covered
by tests, it it not necessary to run the techniques. This happens in
evolution cycles ``+3'' and ``+5''.


We generated \twise{} configurations with the ACTS
tool~\cite{borazjany-etal-icst2012}.  \evopairwise{} and \evosixwise{}
work similarly to \tname{} but note that, as black-box techniques, the
number of configurations they consider is fixed. As \tname{}, they
initially map the functions covered by each test on any reachable
configuration (we used \SPLKorat{} to obtain this information).  Then,
if any of the functions covered by a given test changes,
\evopairwise{} and \evosixwise{} re-run that test against \emph{all
  precomputed configurations}. Otherwise, they discard the test.

\begin{center}
  \fbox{
    \begin{minipage}{8cm}
      \textit{Answering RQ4:}~Perhaps as expected, results indicate
      that \evopairwise{} and \evosixwise{} offer, respectively, lower
      and upper bounds of comparison~--~\tname{} performed in between
      these two extremes.  It ran four times faster than \evosixwise{}
      and ten times slower than \evopairwise{}. Relative performance of
      techniques with respect to the number of configurations is
      similar.
      \end{minipage}
    }
\end{center}

\subsubsection{Effectiveness Results} We evaluated the techniques on 
the ability to detect real bugs. We used the release 4.8.2 of GCC
where the authors of this paper already found bugs\footnote{All bugs
  were confirmed by the GCC team:
  \url{https://gcc.gnu.org/bugzilla/show_bug.cgi?id=<x>}, where
  x=61980, 62069, 62070, 62140, 62141.}~\cite{sabrina-etal-splc15},
and one next release with bugs fixed.  We simulated regression by
``inverting'' the version history: we considered the fixed release as
the initial version (without bugs) and the buggy version as the
subsequent version.  Similar methodology has been used in fault
localization research~\cite{campos-etal-ase2013}.

In this setup, we used 29 tests that we knew a priori would reveal 
crashes in specific configurations. Overall, these crashes exposed 
five distinct bugs. On a crash, the \gcc{} testing infrastructure 
reports an \emph{``Internal Compiler  Error (ICE)''} message followed 
by an specific error description which includes the statement that 
manifested the crash.  We used these messages to identify the crash.


\begin{table}[h!]
\vspace{-1ex}
  \centering  
    \small
    \setlength{\tabcolsep}{5pt}    
    \begin{tabular}{cccccc}
      \toprule
      \multirow{2}{*}{\emph{Technique}} & \multicolumn{5}{c}{\emph{Bug Id}} \\
      \cmidrule{2-6}
       & 1 & 2 & 3 & 4 & 5 \\
      \midrule
      \tname{}       & \cmark & \cmark & \cmark & \cmark & \cmark \\
      \evopairwise{} & \cmark & \cmark & \xmark & \cmark & \xmark \\
      \evosixwise{}  & \cmark & \cmark & \cmark & \cmark & \cmark \\
      \bottomrule
    \end{tabular}
  \vspace{-1ex}
  \caption{\label{tab:effectiveness}Bugs detected in \gcc{} version 4.8.2.}
  \vspace{-1ex}
\end{table}  


Figure~\ref{tab:effectiveness} shows results. Overall, we observed
that \tname{} and \evosixwise{} detected all bugs, while
\evopairwise{} could only reveal three bugs.  The reason is that the
two non-detected bugs are only manifested when a single buggy feature
is enabled (and the others are disabled), and the configurations
generated for pair-wise did not explore this kind of combination with
this specific feature variable.

\begin{center}
  \fbox{
    \begin{minipage}{8cm}
      \textit{Answering RQ5:}~Results indicate that \evopairwise{}
      missed two bugs that alternative techniques found.  \tname{} was
      as effective as \evosixwise{} but it could find bugs much
      faster.
      \end{minipage}
    }
\end{center}


\subsection{Discussion}


We evaluated \tname{} in two scenarios of use: \rcs{} (Regression
Configuration Selection) and \rts{} (Regression Test Selection). The
goals of \rcs{} and \rts{} are the same: to reduce execution cost.
However, the mechanisms to achieve the goal vary. \rcs{} focuses on
reduction of configurations explored by each test (with test set
fixed) whereas \rts{} focuses on reduction of the test set (with
configurations reached in each test fixed).

Compared to our original technique \SPLKorat{} we found that \tname{}
was beneficial in both scenarios, leading to significant reduction of
space and/or time.  Compared to sampling techniques, which are very
popular for testing configurable
systems~\cite{cohen-etal-icse2003,kuhn-etal-tse-2004,kuhn-etal-nist2010,nie-leung-acm-surveys2011},
\tname{} achieved a balance between time and efficiency.  It is
important to note that it is not always clear what best cardinality to
use for combinatorial testing.  Considering the \rts{} scenario, we
observed that, although \evopairwise{} was by far the most efficient
technique, it missed two bugs whereas both \tname{} and \evosixwise{}
caught all bugs in more time. \tname{}, however, detected all bugs
four times faster than \evosixwise{}, on average.

Although performance largely depends on the code under analysis and
code changes, overall, results indicate that our approach is
lightweight and effective.  \tname{} addresses an important gap that
exists between two important fields of research and practice:
configurable system's testing and regression testing.

\section{Threats to Validity}



The main threats to validity are as follows.

\vspace{1ex}\noindent\textbf{External Validity:}~The selection of
subjects used in the experiment with software product lines is not
representative of all programs.  Although open-source and
previously-used in different experiments in the same field, these
subjects are small and may be tested differently compared to real
software.  To mitigate this threat we also considered one real
configurable system that has been under active development for
decades: \gcc{}.  We found that a system like \gcc{} is tested
differently and needs adaptation of a technique like \tname{}.
Another threat is related to the setup we used in our \gcc{}
experiments.  To mitigate this threat we selected tests, options, and
constraints according to a well-defined criteria (see
Section~\ref{gcc:setup}).

\vspace{1ex}\noindent\textbf{Internal Validity:}~Coding errors could
invalidate our results.  To mitigate this threat, we thoroughly
checked our implementation and our experimental results, looking for
discrepancies signaling potential errors

\section{Related Work}
\label{sec:related}

This section presents work most related to \tname{}.


\subsection{Regression Testing for Non-Configurable Systems}
\label{sec:non-conf}


Regression testing is a field of research that can directly impact
in-house testing.  One case that shows such impact in industry is
Ekstazi~\citep{ekstazi-apache}, a recently-available Java library for
regression testing that is used by Apache Camel, CXF, and Commons
Math.  Regression testing also continues to attract attention from the
research community.  Considering regression testing for
non-configurable systems, recent research has focused on the the
following key problems: test selection (\eg{},
~\citep{orso-woda2008,nanda-icst2011,gligoric-etal-issta2015,elbaum-etal-fse2014}),
test-suite prioritization (\eg{}, ~\citep{rothermel-etal-tse2001,
  Saha-icse2015}), test-suite augmentation (\eg{},
~\citep{santelices-ase08,taneja-issat2011,bohme-fse2013}), and
test-suite reduction (\eg{},
~\citep{shi-fse2014,bell-kaiser-icse2014}).  In the following, we
discuss some of these works.

Saha~\etal~\citep{Saha-icse2015} introduce a new approach to address 
the problem of regression test prioritization by reducing it to a 
standard Information Retrieval problem, such that it does not require 
any dynamic profiling or static program analysis to calculate the 
differences between two program versions. 
Elbaum~\etal~\citep{elbaum-etal-fse2014} present two new techniques
for selecting and prioritizing regression tests in the context of
continuous integration development environments.  They use readily
available test suite execution history data to determine what tests
are worth executing and the priority of execution.
Nanda~\etal~\citep{nanda-icst2011} propose an approach to select
regression tests in systems that contain frequent changes in non-code
artifacts. This technique focuses on changes to property files and
databases, complementing code-centric approaches.
Bohme~\etal~\citep{bohme-fse2013} introduce a new regression test
generation technique that stress code where change interaction may
occur, with the purpose of finding more errors. 

None of these works consider the dimension variability to improve
regression testing.  This dimension is very important in many cases
and account for many errors in large real
software~\citep{zhang-ernst-icse2013,
  zhang-ernst-icse2014,medeiros-etal-icse2016}.  It remains open to
explore how to combine these ideas with \tname{} in the context of
configurable systems.


\subsection{Regression Testing for Configurable Systems}
\label{sec:conf}


Previous techniques to solve this problem apply heuristics to find
configurations related to evolutionary
changes~\citep{Qu-etal-icsm2007,Qu-issta2008,Qu-etal-icsm2012,Xu-etal-splc2013}.
We shortly summarize and related some of these works in the following.
Qu \etal{}~\citep{Qu-etal-icsm2007, Qu-issta2008} focus on regression
testing of evolving configurable software systems. They present an
empirical study about the impact of configuration selection heuristics
used in regression testing on fault-detection capability. Their
results highlight that a number of bugs may be missed if certain
configurations are not tested and that prioritizing configurations
allows for more effective testing.  Qu
\etal{}~\citep{Qu-etal-icsm2012} uses slicing-based code change impact
analysis to assist configuration selection.  The technique consists of
two steps.  First, it looks for a set of variables to be re-tested
from a static forward slice of the program.  Then, from this set of
variables, they use pair-wise CIT (Combinatorial Interaction Testing)
to select configurations to be re-tested.  The cost of running the
impact analysis was high but experimental results showed promise when
compared with random and exhaustive selection.

It is important to note that these techniques inherit the limitations
of combinatorial testing: relevant configurations can be missed and
irrelevant configurations can be captured.  \tname{} discovers an
approximation for the set of configurations that require
execution\Comment{ by using a lightweight change-impact analysis}.  It
is very important to note that if one knew a priori that evolutionary
changes would definitely \emph{not} affect the decision tree as
obtained from the previous run of the technique, then it would be
possible to safely re-execute the same configurations from the
previous run without tooling support (and associated
overhead). However, knowing in advance that the structure of the tree
would be preserved requires more sophisticated analysis, which are
often too expensive in this
context~\cite{law-rothermel-icse2003,apiwa-etal-icse05,spllift,lillack-ase2014,angerer-ase2015}.



\subsection{Other}

Zhang and Ernst~\citep{zhang-ernst-icse2013, zhang-ernst-icse2014}
propose a technique to troubleshoot configuration errors caused by
configurable systems' evolution. They use dynamic profiling, execution
trace comparison, and static analysis to link the undesired behavior
to its root cause, a configuration option whose value can be changed
to produce desired behavior from the new software version. This work
starts from failures to find out what changes caused them.  Diagnosis
and repairing faults in configurable systems is an important problem.
It remains to observe how \tname{} could assist in better diagnosis of
configuration problems (\ie{}, in finding root causes of configuration
problems).



The authors of this paper recently found~\citep{souto-etal-icse2017}
that \SPLKorat{} can be combined with sampling to balance cost of
exploration with fault-detection ability.  As future work, we plan to
apply the ideas from that work to improve regression testing of
configurable systems even further.



\section{Conclusions}

Testing configurable systems is important and expensive.  This paper
presented a technique named \tname{} to alleviate cost of systematic
testing these systems when evolutionary information is available.  The
key insight is that it is possible to use simple lightweight impact
analysis to discard configurations and tests during an evolution
cycle.

\tname makes a step forward in closing the gap between research and
practice in testing configurable systems.  We evaluated our technique
on software product lines and on a large configurable systems.  In
both scenarios, results obtained were encouraging.  A prototype
implementation of our technique is available at
\url{https://sites.google.com/site/evosplat/}.

\vspace{1ex}
\noindent\textbf{Acknowledgments.~}Sabrina was supported by the FACEPE
fellowship BPG-0675-1.03/09.  This material is based upon work
partially supported by the Brazilian CNPq research agency under Grant
No. 457756/2014-4 and by a Microsoft SEIF'13 award.

\balance
\bibliographystyle{elsarticle-num}
\bibliography{short-description-testing}

\end{document}